\documentclass[sigconf,nonacm]{acmart}
\AtBeginDocument{%
  \providecommand\BibTeX{{%
    \normalfont B\kern-0.5em{\scshape i\kern-0.25em b}\kern-0.8em\TeX}}}

\setcopyright{acmcopyright}
\copyrightyear{2023}
\acmYear{2023}
\acmDOI{XXXXXXX.XXXXXXX}

\acmConference[CIKM'23]{32nd ACM International Conference on Information and Knowledge Management}{21-25 October, 2023}{Birmingham, UK}
\acmBooktitle{CIKM'23: 32nd ACM International Conference on Information and Knowledge Management, 21th - 25th October, 2023} 
\acmPrice{15.00}
\acmISBN{978-1-4503-XXXX-X/18/06}

\usepackage{wrapfig}
\usepackage{multirow}
\usepackage[british]{babel}

\begin{document}

\title{SC-Block: Supervised Contrastive Blocking within Entity Resolution Pipelines}

\author{Alexander Brinkmann}
\email{alexander.brinkmann@uni-mannheim.de}
\affiliation{%
  \institution{University of Mannheim}
  \city{Mannheim}
  \country{Germany}
}

\author{Roee Shraga}
\email{r.shraga@northeastern.edu}
\affiliation{%
  \institution{Northeastern University}
  \city{Boston}
  \state{MA}
  \country{USA}
}

\author{Christian Bizer}
\email{christian.bizer@uni-mannheim.de}
\affiliation{%
  \institution{University of Mannheim}
  \city{Mannheim}
  \country{Germany}
}

\renewcommand{\shortauthors}{Brinkmann et al.}

\sloppy

\begin{abstract}
The goal of entity resolution is to identify records in different data sources that describe the same real-world entity. However, comparing all records across data sources can be computationally infeasible due to the quadratic runtime complexity of the matching process.
Thus, entity resolution pipelines consist of two parts: a blocker that applies a computationally cheap method to select candidate record pairs and a matcher that afterwards identifies matching pairs from this set using a more expensive method.
This paper presents SC-Block, a blocking method that combines supervised contrastive learning for positioning records in an embedding space with nearest neighbour search for candidate set generation.
We benchmark SC-Block against eight state-of-the-art blocking methods. In order to relate the training time of SC-Block to the reduction of the overall runtime of the entity resolution pipeline, we combine SC-Block with four state-of-the-art matching methods into complete pipelines.
For measuring the overall runtime, we determine the candidate sets that need to be passed to the matcher for maintaining 99.5\% recall.
Our experiments show that SC-Block generates candidate sets that are on average half the size of the candidate sets generated by other blockers. Pipelines using SC-Block execute 1.5 to 2 times faster than pipelines using other blockers, without sacrificing F1 score.
Blockers are often evaluated using relatively small datasets. This might lead to negative runtime effects resulting from large vocabulary sizes being overlooked. In order to measure runtimes in a more challenging setting, we introduce a new benchmark that requires large numbers of product offers from different websites to be blocked. The new benchmark features a maximal Cartesian product of 200 billion pairs.
On the largest dataset of this benchmark, pipelines utilizing SC-Block and the best-performing matcher execute 4 times faster than pipelines utilizing other blockers and the same matcher, reducing the runtime from 30 hours to 8 hours. This runtime reduction clearly compensates for the training time of SC-Block, which in this case is 5 minutes. 
\end{abstract}






\maketitle

\section{Introduction}

Blocking is a crucial step in any entity resolution pipeline because a pair-wise comparison of all records across two data sources is infeasible~\cite{christen_data_2012, thirumuruganathan_deep_2021, papadakis_blocking_2021}.
Blocking applies a computationally cheap method to generate a smaller set of candidate record pairs reducing the workload of the matcher.
During matching a more expensive pair-wise matcher generates a final set of matching record pairs.

In this paper, we propose SC-Block, a supervised contrastive blocking method which applies supervised contrastive learning to position records describing the same real-world entity close to each other in an embedding space.
Supervised blocking reuses the training data, which is commonly used to train a matcher~\cite{li_deep_2020, konda_magellan_2016,peeters_supervised_2022}, to train a blocker~\cite{bilenko_adaptive_2006, zhang_autoblock_2020,papadakis_supervised_2014}.
SC-Block utilizes a supervised contrastive loss that has been successfully applied to computer vision~\cite{khosla_supervised_2020}, information retrieval~\cite{gao_simcse_2021} and especially, matching tasks~\cite{peeters_supervised_2022}.
To illustrate SC-Block's ability to maintain accurate candidate sets we benchmark SC-Block against eight state-of-the-art unsupervised~\cite{christen_survey_2012}, self-supervised~\cite{thirumuruganathan_deep_2021, wang_sudowoodo_2022, mugeni_graph-based_2022} and supervised blockers~\cite{zhang_autoblock_2020, papadakis_supervised_2014}. 

The supervised blocker must not harm the performance of an entity resolution pipeline and the training time of the blocker must be compensated by reducing the runtime of the pipeline.
Therefore, we measure SC-Block's impact on the F1 score of entity resolution pipelines and relate the training time to the pipelines' runtime.
We build the entity resolution pipelines using the mentioned blockers and four state-of-the-art matchers~\cite{konda_magellan_2016,brunner_entity_2020,li_deep_2020,peeters_supervised_2022}.

On three entity resolution benchmark datasets from related work~\cite{konda_magellan_2016,li_deep_2020, thirumuruganathan_deep_2021, wang_sudowoodo_2022}, SC-Block creates the smallest candidate sets and pipelines with SC-Block run 1.5 to 2 times faster than the benchmarked blockers without affecting the F1 score of the pipeline.
The datasets from related work are rather small, which may lead to overlooking runtime effects resulting from many records and many unique tokens generated by tokenizing the record descriptions on whitespace.
Both, many records and a large vocabulary of unique tokens can cause a blocker to build significantly larger candidate sets that increase the workload of the matcher.
To measure the blocker's performance in a more challenging setting, we introduce three new benchmark datasets featuring 200 billion potential comparisons (up to 4,000 times more comparisons than existing benchmarks) and almost 7 million unique tokens within the record descriptions.
On this large benchmark, pipelines utilizing SC-Block and the best-performing matcher execute 4 times faster than pipelines utilizing another blocker with the same matcher reducing the runtime from 30 to 8 hours, clearly compensating for the 5 minutes that are required for training SC-Block. 

\begin{figure*}[t]
\includegraphics[width=.95\textwidth]{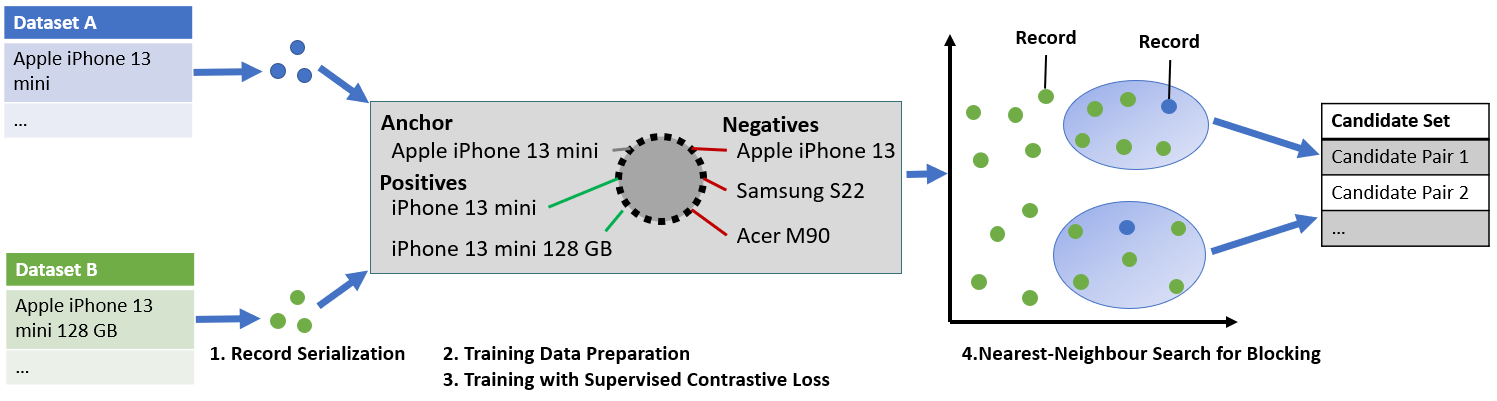}
\caption{The architecture of SC-Block shows how records from the two datasets are embedded (1) and the training data is prepared (2) such that the supervised contrastive loss can position record embeddings describing the same real-world entity close to each other (3) for the nearest-neighbour search of SC-Block (4).}
\label{fig:sc_block_architecture}
\end{figure*}

In this paper, we make the following contributions:
\begin{itemize}
  \item  We propose SC-Block a supervised blocking method which applies supervised contrastive learning to position records describing the same real-world entity close to each other in an embedding space.
  \item We introduce a new large blocking benchmark containing real-world datasets with up to 200 billion record comparisons (4,000 times more comparisons than existing benchmarks) and almost 7 million unique tokens within the record descriptions.
    \item We show that SC-Block creates smaller candidate sets than state-of-the-art blocking methods leading to pipelines that execute 1.5 to 2 times faster on the smaller benchmark datasets and 4 times faster on the new large product matching dataset without negatively affecting F1 scores.
 \item We show that due to the runtime reduction of entity resolution pipelines with SC-Block, it is reasonable to invest training time for SC-Block if training data is available, the number of unique tokens is above 67k unique tokens and the datasets contain at least 10k records to be blocked.
\end{itemize}

The code of all experiments and the new benchmark are available in the project's GitHub repository
\footnote{https://github.com/wbsg-uni-mannheim/sc-block}
.
The paper is structured as follows. In Section \ref{sec:problem_definition} entity resolution is introduced. 
Section \ref{sec:sc_block} presents SC-Block including the supervised contrastive training. In Section \ref{sec:datasets} the datasets from related work including their limitations are discussed. Our new blocking benchmark WDC-B is then presented in Section \ref{sec:wdc_b}. 
The blockers and the impact of the blockers on entity resolution pipelines are evaluated in Section \ref{sec:evaluation_blocking} and in Section \ref{sec:evaluation_pipeline}, respectively. In Section \ref{sec:related_work} related work is discussed.


\section{Problem Definition}
\label{sec:problem_definition}


In this section, we define entity resolution and relate blocking to it. Given two datasets $A$ and $B$, entity resolution aims to find matching records 
that describe the same real-world entity.
The problem of entity resolution can be formalized as follows~\cite{fellegi_theory_1969}:
\begin{definition}[Entity Resolution]
  Given two sets of records $A$ and $B$ and a relation $\sim$, find a subset over all pairs of records $A\times B$ for which $M = \{(r_a,r_b): r_a \sim r_b, r_a \in A, r_b \in B\}$. Similarly, a set of all pairs for which $\sim$ does not hold:
  $U=(A\times B) \backslash M$.
\end{definition}

The relation $\sim$ relates all records representing the same real-world entity. In our scenario, similar to recent works~\cite{zhang_autoblock_2020,papadakis_supervised_2014,thirumuruganathan_deep_2021}, a subset of $M$ and $U$ is known before applying entity resolution that can be used for training and evaluation.

Comparing all record pairs $A\times B$ is computationally infeasible. Therefore, entity resolution is tackled in two steps \emph{blocking} and \emph{matching}. Blocking receives as input all records in $A \cup B$ and returns as output a set of blocks, grouping together similar records while keeping apart dissimilar ones. A set of candidate record pairs $C\subseteq A\times B$ is derived from these blocks. By comparing only similar candidate record pairs, the computational cost of matching is reduced to the comparison of all candidate record pairs instead of $A\times B$. The goal of blocking is to put all matching record pairs into similar blocks while minimizing the number of candidate record pairs. Minimizing the number of candidate record pairs implies that many record pairs are discarded including matching pairs. Therefore, blocking has to find a good trade-off between discarding candidate record pairs and missing matching record pairs~\cite{christophides_overview_2021, christen_data_2012}. In the next section, we outline how SC-Block deals with this trade-off.

\section{SC-Block}
\label{sec:sc_block}

In this section, we introduce a supervised contrastive blocking method named SC-Block. SC-Block applies supervised contrastive learning to position record embeddings describing the same real-world entity close to each other in an embedding space.
Figure \ref{fig:sc_block_architecture} illustrates how the four steps (1) record serialization, (2) training data preparation, (3) training using supervised contrastive learning and (4) nearest neighbour search work together to generate a candidate set. In the following, we introduce the steps in detail and explain how they interact.

\subsection{Record Serialization}
\label{sub-sec:serialization}

Following related work and using the same textual benchmark datasets, SC-Block works with texts~\cite{li_deep_2020, wang_sudowoodo_2022, thirumuruganathan_deep_2021}. 
Therefore, dataset A and B records are serialized into record texts. 
Specifically, let $r\in A\cup B$ be a record, which is composed of multiple attribute values, the serialization of the record is given by $t = Serial(r)$.
For record serialization, we follow a serialization scheme introduced by Li et al.~\cite{li_deep_2020} for entity matching. 
In this serialization scheme, each attribute of a record is serialized as follows, 
"[Col] attribute\_name [Val] actual\_attribute\_value".
For the final record text $t$, all attribute serializations are concatenated.
We refer to a serialized record as record text for the remainder of the paper.

\subsection{Training Data Preparation}
\label{sub-sec:training_data}

We now explain how the training data is prepared for the subsequent supervised contrastive learning step and how source-aware sampling is applied to reduce inter-label noise in the prepared training dataset.
The supervised contrastive loss expects records to share the same label if they describe the same real-world entity.
The input datasets used in this work provide record pairs instead of assigning a label to matching records.
During training data preparation the pairwise matching information from the training and validation set is used to label all records of the input datasets A and B such that matching records share the same label.
For this labelling, we build a correspondence graph, similar to Peeters and Bizer~\cite{peeters_supervised_2022}. The vertices of the correspondence graph are the records, while an edge between two vertices shows that two records match according to the training and validation set. We can then assign a unique label to each connected component of the vertices in the graph so that matching records share the same label.
Step 2 in Figure~\ref{fig:sc_block_architecture} shows how the training data looks after labelling. The left-hand side of the box shows records that share the same label (Positives). The right-hand side of the box shows records that have different labels (Negatives).

\vspace{.1cm}\noindent\textbf{Source-aware sampling.}
The previously introduced labelling can lead to inter-label noise because we only know a subset of matches between the sources and some unlabeled matching records will not share a unique label.
During training, these unlabeled matching records are treated as non-match because they do not share the same label and are not embedded into a close-by location in the embedding space if they appear in the same batch. 
This inter-label noise can reduce the effectiveness of the embeddings~\cite{chuang_debiased_2020}.
To reduce inter-label noise, source-aware sampling is applied~\cite{peeters_supervised_2022}.
Instead of generating one training dataset containing all records and their labels from each input dataset, we generate one training dataset for each input dataset A or B.
Each training dataset contains all records from the respective dataset and only those records from the other table, which share a unique label with a record from the current table.
The training dataset A contains all records from dataset A and all records from dataset B sharing a unique label with a record from dataset A.
The training dataset B contains all records from dataset B and all records from dataset A sharing a unique label with a record from dataset B.
During training, we sample offers from either training dataset A or B into batches. This source-aware sampling strategy eliminates noise due to missing label information, assuming that datasets A and B do not contain duplicates~\cite{peeters_supervised_2022}.

\subsection{Training with Supervised Contrastive Loss}
\label{sub-sec:supervised_contrastive_loss}
In this section, we introduce supervised contrastive loss and how it is applied to train effective embeddings for the nearest neighbour search of SC-Block.
The starting point of the training procedure is a batch of N record texts $t$ sampled from the prepared training data as discussed in Section \ref{sub-sec:training_data}.
We duplicate all records in the batch such that for each record at least one matching record is contained in the batch.
An \textit{encoder network} $Enc(\cdot)$ maps each
record text $t$ to an embedding, $z = Enc(t) \in R$. 
In our experiments, the network encoder $Enc(\cdot)$ is a pre-trained roberta-base transformer model~\cite{liu_roberta_2019} provided by Huggingface\footnote{https://huggingface.co/roberta-base}.
The record embedding $z$ is mean pooled and normalized using the $L_2$ normalization. 
During training, we apply the supervised contrastive loss to update the parameters of the roberta-base model and effectively position the record embedding in the embedding space.
Step 3 in Figure \ref{fig:sc_block_architecture} shows how the supervised contrastive loss exploits label information of matching records by maximizing the agreement of records with the same label (Positives) and minimizing the agreement of records from different labels (Negatives) to train effective embeddings for SC-Block.
We formally define the supervised contrastive loss as follows~\cite{khosla_supervised_2020}.
Given a batch of $2N$ embedded records $z = Enc(t) \in R$:
\begin{equation}
    \mathcal{L} = \sum_{i \in I} \mathcal{L}_i = \sum_{i \in I} \frac{-1}{|P(i)|} \sum_{p \in P(i)} \log \frac{exp(z_i \cdot z_p/\mathcal{T})}{\sum_{a \in A(i)} exp(z_i \cdot z_a/\mathcal{T})}
\end{equation}
Here, $i \in I \equiv {1...N }$ is the index of an embedded record text $z$. 
The index $i$ is called an anchor embedding $z_i$, $P(i) \equiv \{p \in A(i): y_p = y_i\}$ is the set of indices of all records with the same label in the batch distinct from $i$, and $|P(i)|$ is its cardinality. Also recall that $z = Enc(t) \in R$, the dot ($\cdot$) symbol denotes the inner product, $A(i) \equiv I \textbackslash {i}$, and $\mathcal{T} \in R^+$ is a scalar temperature parameter. 
Within a batch, each embedding becomes an anchor embedding $z_i$ that is pulled close to all record embeddings from the same unique label while it is pushed away from all record embeddings with different labels. By setting the batch size to 1024 we make sure that each embedding is compared to many other embeddings, which is beneficial for the supervised contrastive loss~\cite{khosla_supervised_2020}. Additionally, in each training epoch, different records are sampled into a batch leading to a variation in the comparison during training.
The high amount of comparisons and the variations in the comparisons are an advantage of the supervised contrastive loss compared to a pair-wise trained loss that only exploits the annotated record pairs from the training set.
In our experiments, $\mathcal{T}$ is set to 0.07 and the encoder network is trained for 20 epochs with a learning rate of $5e-5$.
After training the encoder network is used to embed all record texts from the input datasets A and B. The trained embeddings $z$ of the encoder network are used for the nearest neighbour search introduced in Section \ref{sub-sec:blocking-framework}.

\subsection{Nearest-Neighbour Search for Blocking}
\label{sub-sec:blocking-framework}

The main goal of Sections~\ref{sub-sec:serialization}-\ref{sub-sec:supervised_contrastive_loss} is to generate record embeddings such that record embeddings describing the same real-world entity are close to each other in the embedding space. Using these representations, as illustrated in Figure~\ref{fig:sc_block_architecture}, we now introduce the nearest neighbour search framework of SC-Block. A blocker defined as a nearest neighbour search is a common approach~\cite{thirumuruganathan_deep_2021, zhang_autoblock_2020, wang_sudowoodo_2022}.

For the search, we define dataset A, which contains fewer records than dataset B, to serve as the query table $Q_A$ and dataset B to serve as an index table $I_B$. Following Section~\ref{sub-sec:serialization}, all records are  serialized ($t = Serial(r)$) and embedded ($z = Enc(t) \in R^D$). For illustration, in Figure~\ref{fig:sc_block_architecture} the query records ($r_A \in Q_A$) are denoted in blue and the index records ($r_B \in I_B$) are denoted in green. The embedded index records are indexed using FAISS~\cite{johnson_billion-scale_2021} to allow an efficient search. FAISS determines the space in bytes required to store the index, which is 4 times $|z|$ times $|I_B|$.

Once all records are serialized, and embedded and the records in $I_B$ are indexed,  the retrieval phase begins by iterating over the query records and searching the index. The search ranks candidate records $z_B\in I_B$ based on their cosine similarity to a query record $z_A \in Q_A$. A hyperparameter $k$ determines how many nearest neighbours are retrieved for each query record. For example, setting $k=5$ means that for each query record, five index records are selected, generating five candidate record pairs. For the nearest neighbour search, FAISS performs a merge-sort on $I_B$, which has a serial time of $O(n \log_2 n)$. Through parallelization on multiple processors, FAISS can reduce this serial time to $O(\log_2 n)$ if a sufficient amount of processors is available. In what follows, reverting back to the original records, a record $r_A\in A$ and its $k$ corresponding neighbouring records $r_1, r_2 \dots, r_k, r_i\in B$ build the set of pairs $(r_A, r_1), (z_A, r_2) \dots (r_A, r_k)$ and are added to the candidate set $C$. Note that by selecting the smaller table as query table $Q_A$, the candidate set grows slower with increasing values of $k$ compared to selecting the larger table as query table and a small candidate set is desired.

\section{Benchmark Datasets}
\label{sec:datasets}

\begin{table*}[t]
\caption{Dataset statistics of the benchmarked datasets.}
\label{tab:dataset_statistics}
\begin{tabular}{@{}lrrrrrrrrrr@{}}
\toprule
DS    & A    & B      & Pos. Pairs & Neg. Pairs & Pos. Pairs & Neg. Pairs & Vocabulary & Max. Matching & Cartesian\\
    &     &      & Train & Train & Val./Test & Val./Test & Size & Neighbours in B  & Product\\ \midrule
A-B   & 1,081 & 1,092    & 616              & 5,127             & 206                 & 1,710                & 7,879 & 2         &       1M        \\
A-G   & 1,363 & 3,266  & 699              & 6,175             & 234                 & 2,059               & 7,389  & 2           &   4M          \\
W-A   & 2,554 & 22,074   & 576              & 5,568             & 193                 & 856               & 49,156  & 3           &   56M          \\
WDC-B$_{small}$ & 5,000 & 5,000  & 9,180             & 11,752            & 500                 & 4,000             & 67,294   & 8         &     25M          \\
WDC-B$_{medium}$ & 5,000 & 200,000  & 9,180             & 11,752            & 500                 & 4,000             & 1,174,280   & 8      &     1,000M             \\
WDC-B$_{large}$ & 100,000 & 2,000,000  & 9,180             & 11,752            & 500                 & 4,000             & 6,880,107   & 8         &     200,000M       \\\bottomrule
\end{tabular}
\end{table*}

We now introduce the benchmark datasets from related work used for our experiments. Table~\ref{tab:dataset_statistics} provides the statistics of the datasets. In this section, we give an overview of the limitations of these benchmarks. In the next section, we present a new benchmark, which overcomes the limitations of these existing benchmarks.

For each dataset, two tables $A$ and $B$ and a ground truth set of positive (matching) record pairs $G_M$ and negative (non-matching) record pairs $G_U$ between the two tables is available.
Columns A and B report the number of records per table.
The ground truth set (containing both $G_M$ and $G_U$) is split into a train, a validation \& a test split.
We report the number of positive and negative pairs per split.
The datasets describe product offers with textual attributes like product name and product description.
The vocabulary size of a dataset is determined by the number of tokens in a dataset. To derive the tokens, the record texts of the dataset are split by whitespace.
The last column reports the maximum number of matching neighbours.
This number determines the smallest value of $k$ necessary to find all matching record pairs between $A$ and $B$ during the nearest-neighbour search if $A$ is used as the query table and $B$ is used as the index table.

\vspace{.1cm}\noindent\textbf{Limitations of benchmark datasets from related work.}
The first three datasets, namely, Abt-Buy (A-B), Amazon-Google (A-G) and Walmart-Amazon (W-A), are publicly available\footnote{https://github.com/anhaidgroup/deepmatcher/blob/master/Datasets.md} entity resolution benchmark datasets~\cite{konda_magellan_2016}, which have been used by other works on matching and blocking~\cite{thirumuruganathan_deep_2021, mudgal_deep_2018, li_deep_2020, peeters_dual-objective_2021, peeters_supervised_2022}.
The datasets are rather small with respect to the Cartesian product $A\times B$. On current hardware, it is possible to omit blocking and compare all record pairs at least for the datasets A-B and A-G. Since a large number of records and a large vocabulary can significantly influence the size of the candidate set generated by a blocker, testing blockers only on small datasets is problematic. Specifically, assuming that each token generated by whitespace tokenization of the records texts in $A$ and $B$ represents a blocking key value, a large number of tokens results in more blocks and a large number of records results in more records per block.
Consequently, a significantly larger candidate set is derived if both the number of blocks and the number of records per block grow. This effect is overlooked if a blocker is only tested on small datasets.
Additionally, a small maximum number of matching neighbours in Table B indicates that a small value of $k$ can be sufficient to find the nearest neighbours for records in Table A.
To overview these limitations, we propose WDC-B in the next section.

\section{The WDC-B Benchmark}
\label{sec:wdc_b}
We propose Web Data Commons Block (WDC-B), a new large blocking benchmark, introducing a more challenging setup to evaluate blockers. 
The dataset is publicly available on our webpage\footnote{https://webdatacommons.org/largescaleproductcorpus/wdc-block/}.


\vspace{.1cm}\noindent\textbf{Seed matching benchmark.}
The basis for this benchmark is the large, pair-wise, 80\% corner cases and 20\% random dataset with the half-seen test set from WDC Products: a multi-dimensional Entity Matching Benchmark\footnote{https://webdatacommons.org/largescaleproductcorpus/wdc-products/}~\cite{peeters_wdc_2023}.
We selected the large version to start with a large number of product offers.
The product offers have been extracted from the Common Crawl using schema.org annotations. Among these annotations are product identifiers like MPNs and GTINs that are used to group offers into clusters for the same product and to build the ground truth split into train, validation and test.
The ground truth of the dataset contains 80\% corner cases and 20\% random matching and non-matching pairs.
Negative corner cases are hard to distinguish non-matching record pairs and positive corner cases are difficult to match matching record pairs.
The authors of the original dataset used a simple binary word occurrence after lower-casing, removing tags and punctuation to cluster product offers in the dataset by similarity.
The clustered record pairs are classified as corner cases or non-corner cases based on a random set of similarity scores to reduce the selection bias.
We select the dataset with 80\% corner cases, which was shown to be challenging for matchers~\cite{peeters_wdc_2023}.
According to the benchmark, also products that are not seen during training are challenging.
We select the 50\% seen test set to have a trade-off between product offers that are part of the training set and not seen during training.

\vspace{.1cm}\noindent\textbf{Three sizes of the benchmark.}
The original dataset of the WDC Products benchmark is split into two separate datasets A and B to match the setup of blocking.
We create a small (WDC-B$_{small}$), a medium (WDC-B$_{medium}$) and a large (WDC-B$_{large}$) versions of the benchmark by filling the datasets A and B with different amounts of additional randomly selected offers from the WDC Product Data Corpus V2020\footnote{http://webdatacommons.org/largescaleproductcorpus/v2020/index.html}.
We make sure that the randomly selected records do not match any of the existing records in the datasets to not introduce additional matching pairs.
Through these additional records, we can measure the effect of a significantly increased vocabulary size between 67K and 6.9M tokens and a significantly increased Cartesian product ($A \times B$) between 25M and 200,000M.
The statistics of the dataset show that WDC-B provides a more challenging setup to evaluate blocking and complete entity resolution pipelines than the other three benchmark datasets.

\section{Blocking-only Evaluation}
\label{sec:evaluation_blocking}
 
We first evaluate SC-Block by analyzing the generated candidate sets of SC-Block against the candidate sets of eight state-of-the-art blockers, which 
are introduced in Section \ref{sub-sec:baseline_blockers}. Details about the implementation of all blockers are given in Section \ref{sub-sec:implementation}.
In Section \ref{sub-sec:fix_nearest_neighbours} we analyze all nearest neighbour blockers with a fixed $k=5$ to derive general conclusions regarding recall and precision of the generated candidate sets.
In Section \ref{sub-sec:variable_nearest_neighbours} we tune the k of the nearest neighbour blockers such that the number of missed matching pairs is minimized.

\subsection{Baseline Blockers}
\label{sub-sec:baseline_blockers}
We now introduce the baseline blockers, which we use to benchmark the performance of SC-Block.
Table \ref{tab:blocking_methods} outlines all blockers including SC-Block with characteristics about the blocker's blocking technique, training procedure, supervision and the encoder network. If a characteristic is not applicable this is marked with `-'. The baseline blockers have been selected from recent related work.

\begin{table}[]
\caption{Supervision, blocking technique, training procedure and encoder network of the benchmarked blocking methods. Supervision can be unsupervised (USL), self-supervised (SL) or supervised (SL). Blocking Techniques are blocking key value (BKV) and nearest neighbour search (NNS).}
\label{tab:blocking_methods}
\begin{tabular}{@{}lllll@{}}
\toprule
                      &  Super-      &  Blocking  &  Training      &  Encoder          \\ 
                      &  vision      &  Technique  &  Procedure      &  Network          \\ 
                      \midrule
JedAI~\cite{papadakis_three-dimensional_2020}                 &  USL     &  BKV        &   -             &  -                        \\
BM25~\cite{paulsen_sparkly_2023}                  &  USL     &  NNS        &  -             &  -                        \\
Auto~\cite{thirumuruganathan_deep_2021}          &  SSL  &  NNS        &  autoencoder  &  fasttext                 \\
CTT~\cite{thirumuruganathan_deep_2021}  &  SSL  &  NNS        &  pair-wise     &  fasttext                 \\
BT~\cite{wang_sudowoodo_2022}           &  SSL  &  NNS        &  contrastive   &  roberta \\
SimCLR~\cite{wang_sudowoodo_2022}                &  SSL  &  NNS        &  contrastive   &  roberta \\
SBERT~\cite{reimers_sentence-bert_2019}         &  SL       &  NNS        &  pair-wise     &  roberta \\
SC-Block              &  SL       &  NNS        &  contrastive   &  roberta \\ \bottomrule
\end{tabular}
\end{table}

\vspace{.1cm}\noindent\textbf{JedAI.}
JedAI defines a blocking key value (BKV) to block records that share six-gram blocks and prunes the candidate set by removing pairs with low matching likelihood to obtain a candidate set~\cite{efthymiou_parallel_2017}.

\vspace{.1cm}\noindent\textbf{BM25.} BM25 is a sparse bag-of-words retrieval model. 
It is an unsupervised nearest neighbour blocker that uses a vector space model~\cite{salton_vector_1975} and the BM25 term weighting scheme~\cite{robertson_probabilistic_2009} to compute a similarity score for the nearest neighbour search. Recent work shows the effectiveness of BM25 for blocking~\cite{paulsen_sparkly_2023}.
We evaluate BM25 with whitespace tokenization referred to as BM25 and BM25 with tri-grams referred to as BM25$_3$.

\vspace{.1cm}\noindent\textbf{Autoencoder (Auto) and Cross Tuple Training (CTT).} Auto and CTT use fasttext to embed the tokens of the record texts and aggregate these token embeddings into a single embedding~\cite{thirumuruganathan_deep_2021}. Both, Auto and CTT need to see all tokens of the datasets during training to learn the token embeddings for the dataset. For Auto, the embeddings are sent through an autoencoder. Auto is self-supervised and requires no labelled data during training. The trained encoder embeds the record texts for the nearest neighbour search. For CTT, the embeddings are sent through a Siamese summarizer.
Afterwards, a classifier learns to detect matches based on the element-wise difference of the created embeddings. CTT is trained on synthetically produced training data derived from the two blocked datasets.

\vspace{.1cm}\noindent\textbf{Barlow Twins (BT) and SimCLR.} The usage of BT and SimCLR for blocking is inspired by Sudowoodo~\cite{wang_sudowoodo_2022}.
During self-supervised training of both models, a batch of record texts is built.
Each record is duplicated, and both the original record text and the duplicate are augmented by dropping a random token of the record text.
The record texts are embedded through a pre-trained roberta language model and mean-pooled. For BT, a linear layer projects the embeddings to 4096 dimensions. 
During training, the empirical cross-correlation of the augmented originals and the augmented duplicates is measured and moved close to the identity matrix~\cite{zbontar_barlow_2021}.
SimCLR maximizes the agreement of embeddings representing the two previously augmented records and minimizes the agreement of embeddings representing distinct records within a batch of records~\cite{chen_simple_2020}.


\vspace{.1cm}\noindent\textbf{Sentence-Bert (SBERT)} SBERT is a sentence embedding framework. Using SBERT's framework we train a supervised nearest neighbour blocker. The blocker requires labelled pairs of matching and non-matching records for training. It embeds record text pairs in a siamese fashion using a pre-trained roberta language model and mean-pools the embeddings.
Afterwards, the cosine similarity of both embeddings is calculated, and a mean-squared error loss is used to fine-tune the weights of the language model~\cite{reimers_sentence-bert_2019}.

\begin{table*}[]
\caption{Recall (R) and Precision (P) of the candidate sets generated by all nearest neighbour blockers with $k=5$ on the test sets of the datasets. The highest Recall and Precision values are marked in bold. The last two columns show the average Recall and Precision over all datasets. 'timeout' indicates a timeout after 48h and 'OOM' indicates an out-of-memory error.}
\label{tab:nns_fixed_k}
 \begin{tabular}{@{}l|cc|cc|cc|cc|cc|cc|cc@{}}
\toprule
  &  \multicolumn{2}{c}{A-B}  &  \multicolumn{2}{c}{A-G}  &  \multicolumn{2}{c}{W-A}  &  \multicolumn{2}{c}{WDC-B$_{small}$}  & \multicolumn{2}{c}{WDC-B$_{medium}$}  & \multicolumn{2}{c}{WDC-B$_{large}$}  &   \multicolumn{2}{c}{Average}\\ 
              &  R      &  P     &  R      &  P     &  R      &  P      &  R      &  P      &  R  &   P &  R  &   P &  R  &   P \\ \midrule
SC-Block & \textbf{99.5\%} & 36.2\% & \textbf{97.4\%} & 35.8\% & 95.9\% & \textbf{32.5\%} & \textbf{71.5\%} & \textbf{57.3\%} & \textbf{66.4\%} & \textbf{63.5\%} & \textbf{56.7\%} & \textbf{74.4\%} & \textbf{81.2\%} & \textbf{50.0\%} \\
BM25$_3$  & 98.1\% & 26.9\% & 94.4\% & 31.2\% & \textbf{96.9\%} & 21.4\% & 52.5\% & 38.9\% & 46.6\% & 43.9\% & \multicolumn{2}{c|}{timeout}   & 77.7\% & 32.5\% \\
BM25     & 92.2\% & 26.7\% & 93.6\% & 31.5\% & 95.9\% & 21.6\% & 59.4\% & 39.7\% & 53.8\% & 45.3\% & 41.7\%               & 54.4\%                  & 72.8\% & 36.5\% \\
Auto     & 75.2\% & 31.3\% & 82.5\% & 35.5\% & 81.9\% & 23.2\% & 43.2\% & 39.5\% & 35.8\% & 40.7\% & \multicolumn{2}{c|}{OOM} & 63.7\% & 34.0\% \\
CTT      & 77.2\% & 31.7\% & 81.6\% & 36.2\% & 81.4\% & 23.6\% & 42.6\% & 38.1\% & 34.8\% & 40.6\% & \multicolumn{2}{c|}{OOM} & 63.5\% & 34.0\% \\
SimCLR   & 90.3\% & 43.7\% & 91.9\% & 36.5\% & 91.2\% & 31.9\% & 34.8\% & 39.2\% & 21.1\% & 38.5\% & 2.7\%                & 33.3\%                  & 55.3\% & 37.2\% \\
BT       & 94.7\% & 29.3\% & 90.2\% & 34.3\% & 92.8\% & 24.8\% & 31.6\% & 34.7\% & 21.3\% & 35.6\% & 12.6\%               & 33.1\%                  & 57.2\% & 32.0\% \\
SBERT    & 74.3\% & \textbf{75.0\%} & 29.1\% & \textbf{41.2\%} & 30.6\% & 16.8\% & 45.3\% & 48.9\% & 35.0\% & 55.1\% & 24.4\%               & 56.8\%                  & 39.8\% & 49.0\% \\ 
 \bottomrule
\end{tabular}
\end{table*}

\subsection{Implementation}
\label{sub-sec:implementation}
A Python implementation of all evaluated pipelines, including blockers and matchers, is shared in our GitHub repository.
We used a shared server with 96 $\times$ 3.6 GHz CPU cores, 1024 GB RAM and an NVIDIA RTX A6000 GPU for the experiments.
For implementing BM25 and BM25$_{3}$, we use Elasticsearch\footnote{https://www.elastic.co/what-is/elasticsearch} with default settings, i.e., neither indexing nor search is tuned.
The Elasticsearch instance runs on a local virtual machine with a Debian 11 Linux distribution, 4 $\times$ 2.1 GHz CPU cores, 32 GBs RAM and 512 GB storage.
The nearest neighbour blockers SC-Block, BT, SimCLR and SBERT rely on FAISS\footnote{https://github.com/facebookresearch/faiss} for indexing the dense record embeddings and the nearest neighbour search using cosine similarity~\cite{johnson_billion-scale_2021}. 
For JedAi, we follow the linked tutorial on entity resolution\footnote{https://github.com/AI-team-UoA/pyJedAI/blob/main/tutorials/CleanCleanER.ipynb}. For Auto and CTT, we use the implementation shared by DeepBlocker~\cite{thirumuruganathan_deep_2021}.

\subsection{Results for Fixed k}
\label{sub-sec:fix_nearest_neighbours}

In this section, we analyze all nearest neighbour blockers with a fixed number of nearest neighbours $k=5$.
We use recall also referred to as pair completeness and precision also referred to as pair quality to evaluate the candidate sets with respect to the test sets of the datasets.
By fixing the hyperparameter $k$ differences in recall and precision become visible that are not visible if $k$ is tuned.
In Section \ref{sub-sec:variable_nearest_neighbours} we tune $k$ such that the generated candidate sets of the nearest neighbour blockers exceed a threshold of 99.5\% on the validation set.
$k=5$ is chosen because it allows the blockers to score high recall, especially on the datasets A-B, A-G and W-A, which exhibit a maximum number of matching neighbours smaller than 5. At the same time, some blockers miss matching pairs, because $k=5$ is not sufficiently large, making it possible to see differences.
Table \ref{tab:nns_fixed_k} shows recall and precision of the candidate sets generated by the nearest neighbour blockers on the test sets of datasets as well as their average across the datasets.


On average, it is evident that SC-Block has the best recall and precision scores among the compared blockers.
It can also be seen that the recall of all generated candidate sets decreases from WDC-B$_{small}$ to WDC-B$_{large}$ and confirms that a large vocabulary and a large cartesian product increase the difficulty of a dataset. 
We now compare SC-Block to the different blockers in more detail.

\vspace{.1cm}\noindent\textbf{Unsupervised blockers.}
The unsupervised blockers BM25 and BM25$_3$ have a similar performance to SC-Block with respect to recall and precision on the datasets A-B, A-G and W-A. On WDC-B BM25 and BM25$_3$ miss on average 16.3\% more pairs than SC-Block. An explanation for this drop in performance is the bigger vocabulary of WDC-B, which makes it difficult for the BM25 blockers to identify matching pairs based on the token overlap. 

\vspace{.1cm}\noindent\textbf{Self-supervised blockers.}
Auto and CTT show the worst performance of the dense nearest neighbour-based blockers on the datasets A-B, A-G and W-A. An explanation is that the pre-trained roberta embeddings of the other blockers are more powerful than the pre-trained fasttext embeddings of Auto and CTT.
SimCLR and BT lose between 3\% and 9\% more pairs than SC-Block on the datasets A-B, A-G and W-A with BT being a bit better than SimCLR.
On W-A the difference between SC-Block, SimCLR and BT is the smallest, which can be explained by the comparably small amount of positive training pairs for this dataset. If supervision is not available, the supervised contrastive loss degenerates to SimCLR~\cite{khosla_supervised_2020}.
On WDC-B SimCLR and BT miss on average 45\% and 43\% more pairs than SC-Block.
SC-Block has a clear advantage because it exploits the supervision of matching and non-matching training pairs found in the datasets.
Interestingly, Auto and CTT outperform BT and SimCLR on WDC-B. This can be explained by the fact that Auto and CTT are trained on all records of the dataset while BT and SimCLR are trained only on the records contained in the development set. This shows how vulnerable the self-supervised blockers are to noise from records they did not see during training. SC-Block's recall and precision scores show that it is more robust against this noise. On WDC-B$_{large}$ Auto's and CTT's nearest neighbour search run out-of-memory showing a limitation of this implementation with respect to the size of the blocked datasets.

\vspace{.1cm}\noindent\textbf{Supervised blockers.}
The supervised blocker SBERT demonstrates the worst performance.
This shows that the supervised contrastive loss of SC-Block better utilizes the supervision than SBERT's pair-wise cosine similarity loss.

Setting $k=5$ is an arbitrary choice.
Section \ref{sub-sec:variable_nearest_neighbours} discusses how $k$ can be tuned such that the blockers miss fewer matching pairs.

\subsection{Results for 99.5\% Recall}
\label{sub-sec:variable_nearest_neighbours}

\begin{table*}[ht]
\caption{$k$ per blocker and dataset after tuning $k$ for recall 99.5\% on the respective validation set. Recall (R) and candidate set size (|C|) of all blockers on the test set. The highest recall, as well as the lowest $k$ and |C| values per dataset, are marked in bold. 'timeout' indicates a timeout after 48h and 'OOM' indicates an out-of-memory error.}
\label{tab:block_variable_k}
\begin{tabular}{@{}c|ccc|ccc|ccc|ccc|ccc|ccc}
\toprule
  &  \multicolumn{3}{c}{A-B}  &  \multicolumn{3}{c}{A-G}  &  \multicolumn{3}{c}{W-A}  &  \multicolumn{3}{c}{WDC-B$_{small}$} & \multicolumn{3}{c}{WDC-B$_{medium}$} & \multicolumn{3}{c}{WDC-B$_{large}$} \\  
 &  $k$  &  R         &  |C|       &  $k$  &  R         &  |C|    &  $k$  &  R        &  |C|      &  $k$  &  R          &  |C|    &  $k$ &  R         &  |C| &  $k$  &  R         &  |C|   \\ \midrule
SC-Block    & 5  & 99.5\%  & \textbf{5k}  & 8  & 99.6\%  & \textbf{11k} & 12 & 96.9\% & \textbf{31k}  & 14                   & 93.5\%               & 70k       & 20                   & 91.9\%               & \textbf{100k}      & 50 & 89.5\% & \textbf{5M}  \\
BM25$_3$       & 13 & \textbf{100\%} & 14k & 27 & \textbf{100\%} & 37k & 12 & \textbf{99.0\%} & \textbf{31k}  & 50                   & 94.2\%               & 250k               & 100                  & 94.0\%               & 500k               &                     \multicolumn{3}{c}{timeout} \\
BM25        & 7  & 94.7\%  & 8k  & 29 & 98.7\%  & 40k & 21 & \textbf{99.0\%} & 54k  & 50                   & \textbf{96.9\%}      & 250k               & 100                  & \textbf{97.8\%}               & 500k               &                   200 & \textbf{95.5\%}  & 20M                          \\
JedAI       & -  & 97.1\%  & 13k & -  & 97.9\%  & 20k & -  & 98.5\% & 172k & - & 55.4\% & \textbf{51k} & - & 80.6\% & 561k &                      \multicolumn{3}{c}{timeout}     \\
Auto        & 50 & 97.1\%  & 54k & 50 & 94.9\%  & 68k & 50 & 93.3\% & 128k & 50 & 85.2\% & 250k & 100 & 80.0\% & 500k &                       \multicolumn{3}{c}{OOM}                            \\
CTT         & 50 & 96.6\%  & 54k & 50 & 94.9\%  & 68k & 50 & 92.2\% & 128k & 50 & 85.2\% & 250k  & 100 & 80.0\% & 500k & \multicolumn{3}{c}{OOM} \\
BT & 20 & 98.5\%  & 22k & 30 & 97.9\%  & 41k & 26 & 98.5\% & 66k  & 50                   & 66.6\%               & 250k               & 100                  & 42.6\%               & 500k               &                       200 & 33.9\% & 20M                            \\
SimCLR      & 29 & 95.6\%  & 31k & 30 & 97.9\%  & 41k & 23 & 96.4\% & 59k  & 50                   & 69.5\%               & 250k               & 100                  & 46.0\%               & 500k               &                       200 & 36.1\% & 20M                        \\
SBERT       & 26 & 85.9\%  & 28k & 30 & 50.9\%  & 41k & 50 & 49.2\% & 128k & 50 & 58.7\% & 250k  & 100 & 78.0\% & 500k & 200 & 58.7\% & 20M                             \\
 \bottomrule
\end{tabular}
\end{table*}

In this section, we analyze how a tuned hyperparameter $k$ influences the recall of the nearest neighbour blockers including SC-Block.
Increasing the $k$ of the nearest neighbour blockers increases the probability of finding a matching pair resulting in a higher recall.
Higher values of $k$ produce larger candidate sets because the matcher has to compare more candidate pairs. This prolongs the matching phase of the entity resolution pipeline.
Our main goal in tuning $k$ is to add all matching pairs to the candidate set while keeping the candidate set as small as possible.
To achieve this goal, we evaluate each nearest neighbour blocker with increasing values of $k$ starting at $k=1$ on the validation set.
Once the recall of the candidate set exceeds 99.5\% on the validation set, $k$ is found.
To limit the search space, we set a maximum value of $k=50$ on A-B, A-G, W-A and WDC-B$_{small}$,  $k=100$ on WDC-B$_{medium}$ and $k=200$ on WDC-B$_{large}$.

Table \ref{tab:block_variable_k} reports $k$, recall on the test set and size of the candidate sets. 
The analysis shows that SC-Block can generate pair complete and smaller candidate sets. Next, we discuss SC-Block's performance in comparison to the other blockers.

\vspace{.1cm}\noindent\textbf{Unsupervised Blockers.}
The candidate sets of the unsupervised blockers BM25$_{3}$, BM25 and JedAi are 2 to 3 times larger than SC-Block on the datasets A-B and A-G.
On W-A the candidate set of JedAi is 5 times larger than the candidate set of SC-Block and BM25$_3$.
On WDC-B benchmark datasets BM25$_{3}$, BM25 and JedAi generate candidate sets that are 5 times larger compared to SC-Block.
The large candidate sets of the unsupervised blockers on the WDC-B benchmark datasets are caused by the high amount of corner cases in the validation sets and the large vocabulary of the datasets.
On the WDC-B benchmark BM25's weighting schema and JedAi's pruning of blockers are not sufficient to reach candidate set sizes that are competitive with SC-Block.
JedAi's pruning of blocks negatively affects the recall score on WDC-B$_{small}$ and WDC-B$_{medium}$ and results in a timeout on WDC-B$_{large}$.
BM25$_{3}$ and BM25 deliver recall scores that are on average above 95\% showing that the blockers can achieve high recall scores if $k$ is sufficiently large.


\vspace{.1cm}\noindent\textbf{Self-supervised Blockers.}
The self-supervised blockers BT and SimCLR create candidate sets that are 2 to 6 times larger compared to SC-Block on the small datasets A-B, A-G and W-A.
Auto and CTT do not reach the recall threshold of 99.5\% on the datasets A-B, A-G and W-A and generate the largest candidate sets.
On WDC-B benchmark datasets none of the self-supervised blockers reaches the recall threshold of 99.5\%.
Again Auto and CTT outperform BT and SimCLR by finding on average 25\% more pairs on the datasets WDC-B$_{small}$ and WDC-B$_{medium}$. The reason for this result is that Auto and CTT see all records during training. The Out-of-memory error on WDC-B$_{large}$ is a result of the nearest neighbour search implementation of Auto and CTT.
These results show that the self-supervised blockers are not robust again noise by records they did not see during training.

\vspace{.1cm}\noindent\textbf{Supervised Blockers.}
SBERT overfits the training and validation data because it reaches a high recall score on the validation set but reaches a much lower recall on the test set.
SC-Block better utilizes the supervision than SBERT's mean-squared error loss on the cosine similarity of a training pair.

Overall, we can see that the candidate sets of the blockers mainly differ in size while the recall of most of the nearest neighbour search blockers is close to 1.
In section \ref{sec:evaluation_pipeline} we evaluate how four state-of-the-art matchers deal with the different candidate sets.

\begin{table*}[]
\setlength{\tabcolsep}{3pt}
\caption{Candidate set size, F1 score and runtime in seconds (RT) of the entity resolution pipelines composed of the blockers SC-Block, BM25$_{3}$ and BT and the matchers SupCon, Ditto, Cross Encoder (CE) and Magellan (Mag.). The highest F1 score, the smallest candidate set size and the lowest RT per dataset are marked in bold. timeout indicates a timeout after 48h}
\label{tab:pip_eff}
\begin{tabular}{@{}ll|ccc|ccc|ccc|ccc|ccc|ccc@{}}
\toprule
         &         & \multicolumn{3}{c}{A-B}                        & \multicolumn{3}{c}{A-G}                          & \multicolumn{3}{c}{W-A}  & \multicolumn{3}{c}{WDC-B$_{small}$}           & \multicolumn{3}{c}{WDC-B$_{medium}$}  & \multicolumn{3}{c}{WDC-B$_{large}$}                         \\ 
Blocker  & Matcher & |C| & F1                         & RT                      & |C| & F1                         & RT                      & |C| & F1                         & RT                      & |C|   & F1                         & RT                      & |C|   & F1                         & RT                       & |C|   & F1                         & RT                      \\ \midrule
SC-Block & SupCon   & \textbf{5k} & 92.6\%          & 71          & \textbf{11k} & \textbf{80.3\%} & 133         & \textbf{31k} & 80.8\%          & 355                      & \textbf{70k} & 70.5\%          & 383          & \textbf{100k} & 71.8\%          & \textbf{742} & \textbf{5M} & 71.7\%          & 30.7k          \\
         & Ditto    & \textbf{5k} & 90.7\%          & 185         & \textbf{11k} & 75.7\%          & 336         & \textbf{31k} & 85.6\%          & 754                      & \textbf{70k} & 76.8\% & 918         & \textbf{100k} & \textbf{78.2\%} & 1.5k                             & \textbf{5M} & \textbf{77.5\%} & 66.5k          \\
         & CE       & \textbf{5k} & 80.0\%          & \textbf{57} & \textbf{11k} & 64.2\%          & 101         & \textbf{31k} & 85.6\%          & 303                      & \textbf{70k} & \textbf{77.1\%}          & \textbf{351} & \textbf{100k} & 77.0\%          & 606          & \textbf{5M} & 75.9\%          & \textbf{27.9k} \\
         & Mag. & \textbf{5k} & 51.6\%          & 335         & \textbf{11k} & 57.7\%          & 511         & \textbf{31k} & 68.0\%          & 1.5k & \textbf{70k} & 59.4\%          & 2.1k                             & \textbf{100k} & 61.0\%          & 2198         & \textbf{5M} & 61.4\%          & 46.2k          \\ \midrule
BM25$_3$    & SupCon   & 14k         & \textbf{92.9\%} & 88          & 37k          & 79.7\%          & 467         & \textbf{31k} & 81.7\%          & 490                      & 250k         & 68.9\%          & 2k                               & 500k          & 69.5\%          & 4.2k                             & \multicolumn{3}{c}{timeout}                    \\
         & Ditto    & 14k         & 90.4\%          & 228         & 37k          & 75.5\%          & 941         & \textbf{31k} & \textbf{86.0\%} & 1k  & 250k         & 74.4\%          & 3.6k                             & 500k          & 75.3\%          & 7.9k                             & \multicolumn{3}{c}{timeout}                    \\
         & CE       & 14k         & 79.0\%          & 70          & 37k          & 64.3\%          & 353         & \textbf{31k} & 85.3\%          & 437                      & 250k         & 73.4\%          & 1.8k                             & 500k          & 73.8\%          & 4k                               & \multicolumn{3}{c}{timeout}                    \\
         & Mag. & 14k         & 50.5\%          & 216         & 37k          & 58.1\%          & 555         & \textbf{31k} & 66.9\%          & 1.9k & 250k         & 44.4\%          & 2.4k                             & 500k          & 43.2\%          & 3.7k                             & \multicolumn{3}{c}{timeout}                    \\ \midrule
BT       & SupCon   & 22k         & \textbf{92.9\%} & 125         & 41k          & 79.7\%          & 157         & 66k          & 80.6\%          & 548                      & 250k         & 58.3\%          & 1.5k                             & 500k          & 45.1\%          & 2.9k                             & 20M         & 39.8\%          & 119.6k         \\
         & Ditto    & 22k         & 90.3\%          & 296         & 41k          & 74.7\%          & 442         & 66k          & 85.7\%          & 1k   & 250k         & 63.9\%          & 3k                               & 500k          & 50.3\%          & 6.3k                             & 20M         & 44.0\%          & 246.1k         \\
         & CE       & 22k         & 79.1\%          & 75          & 41k          & 63.8\%          & \textbf{93} & 66k          & 85.0\%          & \textbf{188}             & 250k         & 62.9\%          & 1.4k                             & 500k          & 49.1\%          & 2.6k                             & 20M         & 43.1\%          & 109.8k         \\
         & Mag. & 22k         & 51.3\%          & 212         & 41k          & 56.5\%          & 346         & 66k          & 65.8\%          & 1.2k & 250k         & 41.6\%          & 2.4k                             & 500k          & 36.2\%          & 2.7k                             & 20M         & 32.47\%         & 72.1k              
\\ 
\bottomrule
\end{tabular}
\end{table*}

\section{Evaluation within Entity Resolution Pipelines}
\label{sec:evaluation_pipeline}
We now evaluate how the blockers SC-Block, BM25$_{3}$ and BT perform within entity resolution pipelines and set the training time of the blockers into context with the runtime of the pipelines.
Therefore, the candidate sets generated by the blockers are evaluated by the state-of-the-art matchers Magellan~\cite{konda_magellan_2016},  Cross Encoder~\cite{brunner_entity_2020}, Ditto~\cite{li_deep_2020} and SupCon-Match~\cite{peeters_supervised_2022} to produce a final set of matching pairs. 
In Section \ref{sub_sec:pip_eff_eff}, we measure how the blockers influence the F1 score and the runtime of the respective pipelines.
In Section \ref{sub-sec:impact_of_training_time}, we analyse whether the required training time of the supervised and self-supervised blockers SC-Block and BT is compensated by the reduced runtime of the respective pipelines.

\subsection{F1 Score and Runtime}
\label{sub_sec:pip_eff_eff}

The goal of this analysis is to understand how the blockers (SC-Block, BM25$_3$ and BT) influence the F1 score and the runtime of entity resolution pipelines.
Runtime covers the execution of blocking and matching and omits training times,
which is discussed in Section \ref{sub-sec:impact_of_training_time}.
Table \ref{tab:pip_eff} shows the candidate set size, F1 score and runtime of the pipelines composed of the three blockers SC-Block, BM25$_3$ and BT and the four matchers SupCon, Ditto, Cross Encoder and Magellan on the six benchmark datasets.

\vspace{.1cm}\noindent\textbf{F1 score.}
From the previous Section \ref{sub-sec:variable_nearest_neighbours} we know that most candidate sets have a high recall. From the results in Table \ref{tab:pip_eff} we can see that for the candidate sets with a high recall the F1 scores of the pipeline mainly depend on the matcher. Pipelines with BT on the WDC-B benchmark are an exception because the low recall of the respective candidate set harms the overall F1 scores. On the WDC-B benchmark datasets, we can see that pipelines, which consist of SC-Block and Ditto or Cross Encoder outperform pipelines with SC-Block and SupCon by 6\% F1 score. The reason is likely that Ditto and Cross Encoder learn different patterns than SC-Block and SupCon. Combining SC-Block with Ditto or Cross Encoder adequately combines these different patterns.

\vspace{.1cm}\noindent\textbf{Runtime.}
On the benchmark WDC-B, we can observe that the small candidate sets of SC-Block lead to pipelines that run 4 to 7 times faster compared to pipelines with BM25$_3$ and BT.
In absolute numbers, the runtime of the pipeline consisting of BT and Cross Encoder on the WDC-B$_{large}$ is 30 hours. SC-Block reduces the workload of the Cross Encoder and reduces the pipeline runtime to 8 hours.
On the WDC-B$_{large}$, BM25$_3$ takes a lot of time to produce large candidate sets, which results in the timeout of the respective pipelines after a runtime of two days.

\begin{table*}[]
\setlength{\tabcolsep}{3.5pt}
\caption{Blocker Training Time (BTT), runtime (RT) and blocker training time plus runtime (CT) of pipelines composed of the blockers SC-Block, BM25$_{3}$ and BT and the matchers SupCon, Ditto, Cross Encoder (CE) and Magellan (Mag.). All values are reported in seconds. The shortest RT and CT per dataset are marked in bold. 'timeout' indicates a timeout after 48h. '-' indicates no training is required.}
\label{tab:run_time_training_time}
\begin{tabular}{@{}ll|ccc|ccc|ccc|ccc|ccc|ccc@{}}
\toprule
         &         & \multicolumn{3}{c}{A-B}                        & \multicolumn{3}{c}{A-G}                          & \multicolumn{3}{c}{W-A}  & \multicolumn{3}{c}{WDC-B$_{small}$}           & \multicolumn{3}{c}{WDC-B$_{medium}$}  & \multicolumn{3}{c}{WDC-B$_{large}$}                         \\ 
Blocker  & Matcher & BTT & RT                         & CT                     & BTT & RT   & CT                      & BTT & RT                         & CT                       & BTT & RT                         & CT                      & BTT & RT                         & CT                      & BTT & RT                         & CT                      \\ \midrule
SC-Block             & SupCon   & 227                   & 71          & 299         & 259                   & 133         & 392          & 430                   & 355          & 785          & 343                   & 383          & 726          & 343                   & 742          & 1.1k         & 343 & 30.7k          & 31.1k          \\
                     & Ditto    & 227                   & 185         & 413         & 259                   & 336         & 595          & 430                   & 754          & 1.2k         & 343                   & 918          & 1.3k         & 343                   & 1.5k         & 1.8k         & 343 & 66.5k          & 66.9k          \\
                     & CE       & 227                   & 57          & 284         & 259                   & 101         & 359          & 430                   & 303          & 733          & 343                   & \textbf{351} & \textbf{694} & 343                   & \textbf{606} & \textbf{949} & 343 & \textbf{27.9k} & \textbf{28.2k} \\
                     & Magellan & 227                   & 335         & 563         & 259                   & 511         & 770          & 430                   & 1.5k         & 2k           & 343                   & 2.1k         & 2.5k         & 343                   & 2.2k         & 2.5k         & 343 & 46.3k          & 46.6k          \\ \midrule
BM25$_3$                & SupCon   & - & 88          & 88          & - & 467         & 467          & - & 490          & 490          & - & 1.9k         & 1.9k         & - & 4.2k         & 4.2k         & \multicolumn{3}{c}{timeout}           \\
 & Ditto    & - & 228         & 228         & - & 941         & 941          & - & 1.1k         & 1.1k         & - & 3.6k         & 3.6k         & - & 7.9k         & 7.9k         & \multicolumn{3}{c}{timeout}           \\
                     & CE       & - & \textbf{70} & \textbf{70} & - & 353         & 353          & - & 437          & 437          & - & 1.8k         & 1.8k         & - & 4k           & 4k           & \multicolumn{3}{c}{timeout}           \\
                     & Magellan & - & 216         & 216         & - & 555         & 555          & - & 1.9k         & 1.9k         & - & 2.4k         & 2.4k         & - & 3.7k         & 3.7k         & \multicolumn{3}{c}{timeout}           \\ \midrule
BT                   & SupCon   & 158                   & 125         & 283         & 236                   & 157         & 394          & 130                   & 548          & 678          & 779                   & 1.5k         & 2.3k         & 779                   & 2.9k         & 3.6k         & 779 & 119.6k         & 120.4k         \\
                     & Ditto    & 158                   & 296         & 455         & 236                   & 442         & 679          & 130                   & 1k           & 1.2k         & 779                   & 3.1k         & 3.8k         & 779                   & 6.3k         & 7.1k         & 779 & 246.1k         & 246.9k         \\
 & CE       & 158                   & 75          & 233         & 236                   & \textbf{93} & \textbf{330} & 130                   & \textbf{188} & \textbf{317} & 779                   & 1.4k         & 2.1k         & 779                   & 2.6k         & 3.4k         & 779 & 109.8k         & 110.6k         \\
 & Magellan & 158                   & 212         & 370         & 236                   & 346         & 582          & 130                   & 1.2k         & 1.3k         & 779                   & 2.4k         & 2.4k         & 779                   & 2.7k         & 2.7k         & 779 & 72.1k          & 72.9k    
   \\ 
\bottomrule
\end{tabular}
\end{table*}

On the datasets A-B, A-G and W-A, the smaller candidate sets of SC-Block lead to pipelines that finish 1.5 to 2 times faster compared to pipelines with the other blockers.
Generally, we can conclude that an effective blocker like SC-Block can reduce the overall runtime of a pipeline on small datasets.
To see if training an effective blocker for the different data sets is reasonable, we set the reduction in runtime into context with training time in Section \ref{sub-sec:impact_of_training_time}.

\subsection{Impact of Training Time}
\label{sub-sec:impact_of_training_time}

In this section, we relate the training time of the blockers SC-Block and BT to the overall runtime of the entity resolution pipelines.
Training of a blocker is only reasonable if the runtime of the pipeline including training time is shorter than the runtime of a pipeline with a blocker that is not trained.
In Table \ref{tab:run_time_training_time} we show the blocker training time (BTT), the runtimes (RT) and the sum of training time and runtime (CT) of all pipelines on the six datasets.

\vspace{.1cm}\noindent\textbf{Small Benchmark Datasets.}
For the small datasets, A-B, A-G and W-A supervised and self-supervised blocking methods are not substantially more efficient than the unsupervised BM25$_3$ blocker. On A-B, A-G and W-A, the small gain in runtime through SC-Block and BT compared to BM25$_3$ is withdrawn by the training time of the blockers.
It is not reasonable to train SC-Block for these small datasets, because entity resolution pipelines with the unsupervised BM25$_3$ blocker and a tuned $k$ value produce recall scores close to 1, similar F1 scores and have a shorter runtime compared to SC-Blocks training time and runtime. 

\vspace{.1cm}\noindent\textbf{WDC-B Benchmark.}
On the WDC-B benchmarks, the runtime of pipelines utilizing BM25$_3$ exponentially grows with a larger vocabulary and a larger cartesian product of candidate pairs.
In this case, training SC-Block is reasonable because the training data is available and the five minutes of training time reduce the runtime of the pipeline compared to BM25$_3$ even on WDC-B$_{small}$ by 20 minutes for the matchers SupCon and Cross Encoder.
Given our compute restrictions, running competitive pipelines on WDC-B$_{large}$ was only possible if SC-Block was applied for blocking.

\section{Related Work}
\label{sec:related_work}


Entity resolution is key for data integration. Thus, both blocking and matching have been researched for decades~\cite{christen_data_2012, cohen_learning_2002, konda_magellan_2016, papadakis_blocking_2021}.

\vspace{.05cm}\noindent\textbf{Blocking.}
Blocking is traditionally tackled as an unsupervised task by extracting a blocking key value from each record~\cite{christen_survey_2012, aizawa_fast_2005}. Records with the same blocking key value are assigned to the same block. Meta-blocking extends blocking by blocking key values with an additional pruning step that first weights candidate record pairs by their matching likelihood and discards pairs with the lowest scores~\cite{efthymiou_parallel_2017}. This pruning step has also been implemented as supervised meta-blocking leveraging supervision to discard candidate pairs~\cite{gagliardelli_generalized_2022,papadakis_supervised_2014}.
Related works also utilized supervision to directly learn a blocking strategy~\cite{bilenko_adaptive_2006, zhang_autoblock_2020}.
Recently, deep learning for blocking has become popular. DeepBlock~\cite{javdani_deepblock_2019} was the first framework to use embeddings for blocking. 
AutoBlock~\cite{zhang_autoblock_2020} uses labelled data to position the record embeddings in the embedding space and uses locality-sensitive hashing to retrieve the nearest neighbours for candidate set building. 
DeepBlocker~\cite{thirumuruganathan_deep_2021} extensively explored the use of deep self-supervised learning for blocking, introducing two state-of-the-art blockers based on auto-encoding (Auto) and cross-tuple training (CTT).
Recently, Sudowoodo~\cite{wang_sudowoodo_2022} has been proposed, which applies self-supervised learning in combination with a transformer model and the two loss functions SimCLR and Barlow Twins.
The supervised contrastive loss of SC-Block is an extension of the SimCLR loss.
We compare SC-Block to the blockers meta-blocking through the JedAi toolkit, Auto, CTT, SimCLR and Barlow Twins from the related work.
Except for JedAI all benchmarked blockers apply the nearest neighbour search for candidate set generation.
Mugeni et al. use self-supervised contrastive learning to position embeddings in the embeddings space and apply an unsupervised community detection technique from graph structure mining to block record pairs~\cite{mugeni_graph-based_2022}.
Most recently, Sparkly has demonstrated that TF-IDF and BM25 are strong blockers for entity matching~\cite{paulsen_sparkly_2023}.

\vspace{.05cm}\noindent\textbf{Entity Matching.}
Deep learning became a major player in designing effective entity matching methods, initiated by Ebraheem et al.~\cite{ebraheem_distributed_2018} 
and Mudgal et al.~\cite{mudgal_deep_2018}, mainly focusing on the matching phase. 
Recently, several matching methods that use transformers as pre-trained language models achieve state-of-the-art performance~\cite{li_deep_2020,brunner_entity_2020, peeters_dual-objective_2021}.
Among them, Peeters and Bizer~\cite{peeters_supervised_2022} and Wang, Li and Wang~\cite{wang_sudowoodo_2022} use contrastive learning, similar to our work. 
The state-of-the-art matchers Ditto, Cross Encoder and SupCon are used in our work to compose complete entity resolution pipelines.

\vspace{.05cm}\noindent\textbf{Contrastive Learning.}
SC-Block builds on recent works in Information Retrieval~\cite{gao_simcse_2021}, Computer Vision~\cite{khosla_supervised_2020} and Entity Matching~\cite{peeters_supervised_2022}, for which contrastive learning was shown to be more effective than the traditional cross-entropy-based learning.  
Specifically, Gao, Yao and Chen~\cite{gao_simcse_2021} use contrastive learning to learn sentence embeddings without any supervision. Supervised contrastive learning still suffers from challenges, including the robustness of learned representations~\cite{chuang_debiased_2020} and class collapse, i.e., all samples from a cluster are mapped to the same representation~\cite{chen_perfectly_2022}.
Similar to SupCon~\cite{peeters_supervised_2022}, SC-Block applies source-aware sampling to increase the robustness of the learned embeddings.
The main difference to SupCon is that SC-Block's embeddings are optimized for blocking and not for matching, which requires longer pre-training of the embeddings.


\section{Conclusion}
\label{sec:conclusion}

In this paper, we proposed SC-Block a supervised contrastive blocking method which combines supervised contrastive learning for positioning records in an embedding space and nearest neighbour search for candidate set building. It utilizes training data that is already available for matcher training to train an effective blocker.
We benchmark SC-Block against eight state-of-the-art blockers and within entity resolution pipelines in combination with four state-of-the-art matchers.
On three product-matching datasets from related work, SC-Block creates the smallest candidate sets leading to pipelines that run 1.5 to 2 times faster compared to pipelines with the benchmarked blockers maintaining similar F1 scores.
We further introduce a new benchmark dataset WDC-B, representing a large challenging blocking scenario with up to 200 Billion potential record comparisons.
On WDC-B, the pipeline with SC-Block and the best-performing matcher runs 4 times faster than pipelines with state-of-the-art blockers and the same matcher reducing the runtime from 30 to 8 hours compensating for the required 5 minutes of training time. 

\bibliographystyle{ACM-Reference-Format}
\bibliography{library}

\end{document}